\newif\ifcomm
\commfalse

\newif\ifs
\sfalse

\newif\ifblind
\blindfalse

\newif\ifhyperref
\hyperreftrue

\newif\ifconf
\conffalse

\newcounter{format}
\newif\ifacm
\ifnum \value{format} < 2 \acmtrue \else \acmfalse \fi
\newif\ifacmart %
\ifnum \value{format} = 0 \acmarttrue \else \acmartfalse \fi
\newif\ifusenix
\ifnum \value{format} = 2 \usenixtrue \else \usenixfalse \fi
\newif\ifhotnets
\ifnum \value{format} = 3 \hotnetstrue \else \hotnetsfalse \fi
\newif\ifieee
\ifnum \value{format} = 4 \ieeetrue \else \ieeefalse \fi
\newif\ifnines
\ifnum \value{format} = 5 \ninestrue \else \ninesfalse \fi

\ifconf
    \newcommand{\Conf}[1]{#1}
    \newcommand{\TR}[1]{}
    \newcommand{\Journal}[1]{}  %
    \newcommand{\OnlyTR}[1]{}   %
\else
    \newcommand{\Conf}[1]{}
    \newcommand{\TR}[1]{#1}
    \newcommand{\Journal}[1]{#1}  %
    \newcommand{\OnlyTR}[1]{}   %
\fi

\ifblind
    \documentclass[sigconf, 10pt, anonymous, nonacm]{acmart}
\else
    \documentclass[sigconf, 10pt, authorversion, nonacm, screen]{acmart}
\fi

\ifieee %
    \usepackage{tikz}
    \usepackage[noadjust]{cite} %
	\usepackage[cmex10]{amsmath}
	\usepackage{amsthm}  %
    \newtheoremstyle{boldthm}{}{}{\itshape}{}{\bfseries}{.}{ }{\thmname{#1}\thmnumber{ #2}\thmnote{ (#3)}} %
    \theoremstyle{boldthm}
            
\else
    \usepackage{amsmath}
    \usepackage{amsthm}

\fi

\ifnines
    \usepackage{graphicx}
    \usepackage[tt=false, type1=true]{libertine}
    \usepackage{mathptmx}
    \usepackage{amssymb}
    \usepackage{tikz}
    \usepackage[varqu]{zi4}
    \usepackage[T1]{fontenc}
    \usepackage[font={footnotesize},labelfont={footnotesize,bf},textfont={footnotesize,it}]{caption}
\fi
\ifhotnets
    \usepackage{times}  
\fi

\ifacmart 
\else 
    \ifusenix
    \else
        \ifhotnets 
        \else
            \ifhyperref
                \usepackage[hidelinks]{hyperref} %
                \hypersetup{pdfstartview=FitH,pdfpagelayout=SinglePage}
            \fi
        \fi
    \fi
\fi

\ifacmart
\else
  \newtheorem{theorem}{Theorem}%

\fi

\ifieee %
    \newcommand{\bp}{\begin{IEEEproof}}     %
    \newcommand{\bpo}{ \begin{IEEEproof}[Proof Outline] }
    \newcommand{\ep}{\end{IEEEproof}}       %
    \newcommand{\proofof}[1]{\begin{IEEEproof}[Proof of #1]} %
\else
    \newcommand{\bp}{\begin{proof}}
    \newcommand{\bpo}{ \begin{proof}[Proof Outline] }
    \newcommand{\ep}{\end{proof}}       %
    \newcommand{\proofof}[1]{\begin{proof}[Proof of #1]} %
\fi

\ifieee
    \usepackage[justification=justified, font=footnotesize]{caption}
    \usepackage[labelformat=simple, font=footnotesize]{subcaption}  
    \DeclareCaptionLabelSeparator{periodspace}{.\quad}
    \ifconf
    \else
    \fi
\else %
    \ifnines \else
        \ifs %
            \usepackage[small]{caption} 
        \else
            \usepackage{caption}
        \fi
    \fi %
    \usepackage[labelformat=simple]{subcaption}
\fi

\usepackage{graphicx}
\usepackage{xurl}
\usepackage[normalem]{ulem} %
\usepackage{xcolor} %

\usepackage{xspace}
\usepackage{balance}
\usepackage{booktabs}  %
	
\usepackage{array}
	\newcolumntype{C}[1]{>{\centering\let\newline\\\arraybackslash\hspace{0pt}}m{#1}}

\usepackage[inline,shortlabels]{enumitem} %
\newcommand{\bl}{\begin{enumerate*}[(1)]}
\newcommand{\el}{\end{enumerate*}}

\usepackage[capitalize,noabbrev]{cleveref}
\crefname{equation}{Eq.}{Eqs.}
\Crefname{equation}{Eq.}{Eqs.}
\crefname{figure}{Fig.}{Figs.}
\Crefname{figure}{Fig.}{Figs.}
\crefname{table}{Table}{Tables}
\Crefname{table}{Table}{Tables}
\crefname{property}{Property}{Properties}
\Crefname{property}{Property}{Properties}
\crefformat{section}{\S#2#1#3}
\crefformat{subsection}{\S#2#1#3}
\Crefformat{section}{\S\S#2#1#3} %
\Crefformat{subsection}{\S\S#2#1#3} %

\ifcomm
  \newcommand{\mycomm}[3]{{\footnotesize{{\color{#2} \textbf{[#1: #3]}}}}}
  \newcommand{\Fmycomm}[3]{{\color{red} \footnote{{{\color{#2} \textbf{[#1: #3]}}} }}}
\else
    \newcommand{\mycomm}[3]{}
    \newcommand{\Fmycomm}[3]{}
\fi

\ifs
	\usepackage[compact]{titlesec} %

    \setlist{leftmargin=*} %
    \newcommand{\T}[1]{\par\vspace{2pt plus 1pt minus 1pt}\noindent\textbf{#1}} %
    \newcommand{\Ts}[1]{\par\noindent\textit{#1}} %

\else

    \newcommand{\T}[1]{\par\smallskip\noindent\textbf{#1}} %
    \newcommand{\Ts}[1]{\par\smallskip\noindent\textit{#1}} %
\fi
\newcommand{\mypar}[1]{\T{#1.\xspace}}
\newcommand{\mypars}[1]{\Ts{#1.\xspace}}

\newcommand{\be}{\begin{equation}}
\newcommand{\ee}{\end{equation}}

\newcommand{\floor}[1]{\left\lfloor #1 \right\rfloor} 
\newcommand{\Tmax}{\ensuremath{T_{\max}}} %

\newcommand{\vx}{\checkmark\kern-1.1ex\raisebox{.7ex}{\rotatebox[origin=c]{125}{--}}} %

\usepackage{pifont} %

\providecommand{\ie}{{i.e.,}\xspace}
\providecommand{\eg}{{e.g.,}\xspace}

\newcommand{\newVar}[2]{\newcommand{#1}{\ensuremath{#2}\xspace}}
\newcommand{\renewVar}[2]{\renewcommand{#1}{\ensuremath{#2}\xspace}}
  \newVar{\server}{S}
  \newVar{\client}{C}
  \newVar{\rclient}{R_c}
  \newVar{\rserver}{R_s}
  \renewVar{\th}{\ensuremath{^\text{th}}}

\newcommand{\mySC}[1]{\textsc{#1}\xspace}
\newcommand{\name}{\mySC{CAPS}}

\begin{document}

\title{\name: Fine-Tuning CCA Timing}

\ifblind
\else
    \ifhyperref
        \newcommand{\aut}[2]{#1\texorpdfstring{$^{#2}$}{(#2)}}  %
    \else
            \newcommand{\aut}[2]{#1$^{#2}$}
    \fi
    \author{
          \aut{Raphael Zailer}{1},
          \aut{Isaac Keslassy}{1,2}
        }%
    \affiliation{
        $^1$ \textit{Technion} \quad 
        $^2$ \textit{UC Berkeley}%
    \country{}%
    }
    \renewcommand{\shortauthors}{R. Zailer and I. Keslassy}     %
\fi

\ifacm %
    \sloppypar
\else 
    \ifhotnets
        \sloppypar
    \else
    \fi
\fi

\begin{abstract}
Data-center congestion control targets high throughput, fair bandwidth allocation, and low latency. Modern transports couple rate computation and packet scheduling into a single feedback loop, converging to near-optimal rates but leaving standing queues that can scale with the number of flows. 
We argue that separating the two reveals a simpler design point. Given stable feasible rates, the residual queue problem reduces to a \emph{timing} problem: if every flow's packets arrive at the bottleneck in the correct slot, the link stays busy and the queue stays empty. Clocked ACK-Paced Synchronization (\name) is a lightweight distributed scheduling layer that achieves this by phase-locking each sender's transmissions to ACK-clocked bottleneck slots, with a per-flow correction that compensates for heterogeneous RTTs. We characterize the phase-locked steady state for dumbbell topologies under equal RTT, heterogeneous RTT, and bidirectional traffic, and validate the mechanism on a fat-tree under incast, permutation, and all-to-all traffic. \name reduces worst-case queue occupancy by $5$--$10\times$ across all tested scenarios without throughput loss.
\end{abstract}
\maketitle

\section{Introduction}

A congestion control algorithm manipulates two levers: (1)~\emph{what} to send (the rate or window size), and (2)~\emph{when} to send it (ACK-clocking, pacing, or some other release discipline). The first lever has been studied for decades; modern data-center CCAs (Swift~\cite{swift}, HPCC~\cite{hpcc}, S-PERC~\cite{sperc}) can converge to near-optimal max-min fair rates. The second lever has barely been explored~\cite{tcp-pacing, ack-clock-06, ack-clock-08}. This paper argues it is the key to eliminating persistent queues in long-lived, predictable datacenter workloads, such as bulk transfers and AI/ML collectives where rates are stable and RTTs are known.

Persistent queues are not merely a latency nuisance, they are a scaling bottleneck.  With $N$ unsynchronized periodic flows, queue occupancy typically scales as $\Theta(\sqrt{N})$ (as observed by HPCC~\cite{hpcc} and Swift~\cite{swift}). Unfortunately, switch buffers consume on-chip SRAM that dominates both die area and power dissipation. As we scale the number of flows $N$, the need for deep buffers currently increases just as we reach a point where they become too costly: At 1.6\,Tbps port speeds, deeper buffers multiply both memory cost and energy per bit forwarded. Thus, we deploy many mechanisms to reduce  those queues, such as ECN marking, PFC pause frames, and CCA window back-off, adding protocol complexity and triggering cascading slowdowns. If queues could be kept near-empty by construction, much of this machinery becomes irrelevant.

Existing techniques do not solve the timing problem. Intra-flow pacing smooths micro-bursts from a single sender but cannot coordinate timing \emph{across} flows: two independently paced flows can still collide persistently if their phases drift. ACK-clocking couples a flow's cadence to the bottleneck rhythm, but only when all flows share the same RTT. With heterogeneous RTTs, shorter-RTT flows receive feedback sooner and gradually steal timing phase, rebuilding the queue that pacing was supposed to prevent. No deployed scheme accounts for this inter-flow phase drift.

In this paper, we introduce Clocked ACK-Paced Synchronization (\name). \name exploits the second lever directly. Given converged fixed rates (from any allocator), it decides \emph{when} each packet is released. It abstracts time as a sequence of frames and makes sure that flows reuse the same bottleneck slots that are assigned to them across consecutive frames. To do so, it relies on a simple ACK-clocking with a small internal delay correction.
The result is a phase-locked schedule: after about one RTT, queue occupancy drops to zero while utilization remains at or near 100\%.

Finally, we evaluate CAPS on datacenter topologies under all-to-one (incast), permutation, and all-to-all traffic. We show that CAPS reduces worst-case queue occupancy by $5-10\times$ across all tested scenarios, while keeping the same throughput and relying on a simple lightweight implementation.

\section{Background and Related Work}

\mypar{Convergence to optimal rates} Maintaining small buffers is a primary service-quality requirement in data centers. Small queues reduce tail latency for short RPC-style flows~\cite{dctcp-tail}, mitigate bufferbloat~\cite{bufferbloat}, and limit head-of-line blocking under incast bursts~\cite{incast}. Because links are fast and RTTs are short, even a modest standing queue imposes a meaningful latency penalty.

Modern transports attack queues through continuous feedback. S-PERC computes per-flow max-min fair rates from link-level capacity sharing~\cite{sperc}. Swift uses delay/RTT signals to keep queues near a target~\cite{swift}. HPCC uses in-band telemetry (INT) to estimate real-time utilization and applies rapid rate correction~\cite{hpcc}. All three reach a good approximation in practice to the optimal fair rates.

\mypar{Pacing}
Pacing spaces a sender's packets evenly over an RTT to smooth bursts upon receiving ACKs~\cite{tcp-pacing} and is widely deployed (Linux FQ/pacing, BBR~\cite{bbr}).
As discussed above, it cannot coordinate timing across flows or across successive rounds. A lighter variant, \emph{initial-window pacing} (IWP), paces only the first window of packets and then relies on ACK clocking~\cite{tcp-pacing-dc, rfc6928}.

\mypar{ACK clocking}
Jacobson's packet-conservation principle~\cite{jacobson88} established that ACK-clocked senders naturally match the bottleneck service rate: each returning ACK releases exactly one new packet. Subsequent work modeled ACK-clock dynamics as an inner feedback loop~\cite{ack-clock-06, ack-clock-08} and analyzed stability in single-link settings~\cite{fast-tcp-stability-05}. 

\mypar{Buffer sizing}
Classical buffer sizing rules (e.g., bandwidth-delay product divided by $\sqrt{N}$~\cite{buffer-sizing}) provision enough buffer to absorb statistical traffic fluctuations in the case of TCP-like CCAs. 

\mypar{TDMA scheduling}
Time-division schemes assign flows to fixed time slots, guaranteeing collision-free access~\cite{fastpass}. We share the goal of collision avoidance but intend to avoid any  centralized slot assignment or global clock. Independently, with $N$ flows sending periodically at fair rates, standing queue occupancy can be modeled as scaling as $\Theta(\sqrt{N})$, regardless of how precisely rates converge (as follows from the ND/D/1 queueing model and detailed in  HPCC~\cite{hpcc} and Swift~\cite{swift}). 

\section{CAPS Algorithm}

In this section, we introduce the Clocked ACK-Paced Synchronization (\name) algorithm. We  start by providing intuition in a toy topology, before extending to more general topologies.

\mypar{Dumbbell topology}
We start by considering a standard dumbbell network with $n$ flows, where in each flow $i$ source host $i$ sends an infinite-size flow to destination host $i$. We assume two intermediate switches connected by a bottleneck link of capacity $C$.
The switches have FIFO output queues. All links have the same capacity $C$, but different propagation delays.

In addition, we assume that each flow $i$ knows its fixed propagation RTT $T_i$, as well as the worst-case propagation delay in the network $\Tmax=\max_i T_i$. We also assume that each flow $i$ targets a given constant rate $x_i$ by using a window-based CCA with window size $W_i=\floor{x_i \cdot T_i}$, such that the total rate cannot exceed the bottleneck link capacity: $\sum_{i=1}^{n} x_i < C$. The fixed window assumption is important: We rely below on the windows not changing too fast (\ie, being kept constant for more than an RTT) in order to establish several properties for the queue size.

\mypar{Equal RTTs} Assume at first that all flows have the same RTT. Then our algorithm CAPS simply uses a \textit{standard ACK clocking} mechanism: it always keeps $W_i$ unacknowledged packets on the line, and sends the next packet upon receiving the next ACK. Formally, we wait to receive the ACK for packet $k$ in order to release packet $k + W_i$. 

We find the following result: \textit{ACK clocking zeroes the queue size after an RTT}. This is because after the initial packets are stuck in the bottleneck queue, they leave it in a FIFO order without overlap. The ACKs then keep this order and this global synchronization due to the equal RTTs, and this is later reflected in the packets sent in the next RTT. Thus, \textit{the network reaches a global synchronization with a fully distributed ACK-clocking mechanism}.

\begin{figure}
\centering
\begin{subfigure}{\columnwidth}
  \centering
  \includegraphics[width=0.85\columnwidth]{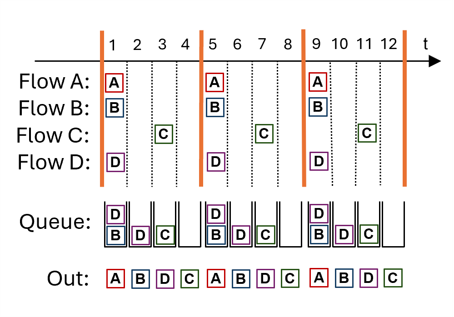}
  \caption{Random periodic arrivals: persistent queue of 2~packets despite stable output.}
  \label{fig:ex1-random}
\end{subfigure}
\begin{subfigure}{0.95\columnwidth}
  \centering
  \includegraphics[width=\columnwidth]{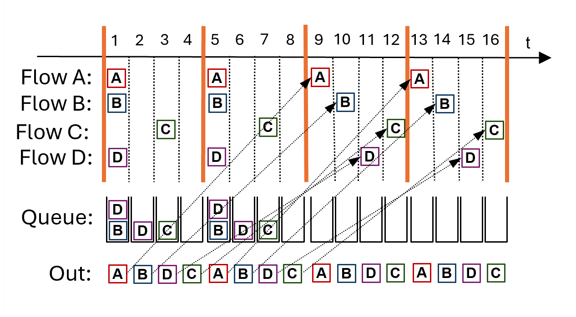}
  \caption{ACK-clocked: queue drains to zero by round~3 while maintaining full utilization.}
  \label{fig:ex1-ack}
\end{subfigure}
\caption{Equal-RTT example with 4 flows ($W_i = 2$ each, RTT = 8 slots). ACK clocking naturally aligns arrivals to bottleneck slots.}
\label{fig:example1}
\end{figure}

\Cref{fig:example1} provides more intuition, assuming that time is slotted to make it easier to visualize. It illustrates four flows (A--D) with $W_i = 2$ and RTT = 8~slots. \Cref{fig:ex1-random} illustrates a simple algorithm where each flow uses IWP, then periodic arrivals without ACK clocking. This simple approach results in a persistent collision and non-zero queues (it can be modeled using a D/D/1 queueing model~\cite{kleinrock75}).
On the contrary, \cref{fig:ex1-ack} shows that CAPS leverages ACK clocking to get all packets globally synchronized and manages to fully fill the link. Specifically, as packets get queued, they are delayed and therefore they also delay their corresponding ACKs. After an RTT, the packets are effectively synchronized, the queue is empty, and we reach 100\% link utilization. 

Formally, we can prove the following result: 
\begin{theorem}
    Using CAPS in a dumbbell topology with equal RTTs, the bottleneck queue size converges to zero within an RTT.   
\end{theorem}
\bpo
    We outline the proof due to space reasons. \textbf{(1)}~Lindley's recurrence at the bottleneck gives departures that are always $\tau$-spaced regardless of arrival collisions. \textbf{(2)}~Because downstream links have equal capacity and fixed delay, this spacing is preserved to the destination and back; no secondary queue forms and ACKs return in sending order. \textbf{(3)}~Under ACK-clocking, each successor packet arrives at the bottleneck exactly one RTT after its predecessor's departure, so the inter-arrival gaps equal the original departure gaps ($\ge\tau$). \textbf{(4)}~The frame-level load constraint ensures the transient backlog clears by the end of the first frame; from round~2 onward every packet finds an empty queue.
\ep

\mypar{Heterogeneous RTTs} 
When all flows have different RTTs, basic ACK clocking cannot globally synchronize all flows. Flows with a shorter RTT receive feedback sooner and aggressively send packets earlier, thus losing the synchronization with longer-RTT flows.

The idea in CAPS is to synchronize all flows as if they all had the same worst-case RTT $\Tmax$. Specifically, we employ the combination of two techniques:

\noindent \textbf{(1)} We use an adjusted window 
\be
W^m_i = \floor{x_i \cdot \Tmax}.  
\ee
Thus, the adjusted window ensures that each flow still sends at approximate rate $x_i$ over the common $T_{\max}$ frame.

\noindent \textbf{(2)} When an ACK on packet $k$ comes back to sender $i$, we delay the transmission of packet $k+W^m_i$ by $\Tmax-T_i$. This is because we simulate the fact that all flows have the same RTT $\Tmax$, so essentially we synchronize back all flows in a distributed way. Note that the delay only shifts the packet departure times, not their sending rate. 

To summarize, \name relies on a \emph{frame-alignment} method, rather than only being an in-flight limiter as in most CCAs. In heterogeneous-RTT settings, all flows are mapped onto a common frame duration (T$_{\max}$) so their combined packet budget matches what the bottleneck can serialize during that frame. \name is easiest to understand as a repeating loop:
\begin{enumerate}[leftmargin=1.6em]
\item The sender transmits a packet.
\item The packet reaches the bottleneck and is serialized there.
\item The receiver sends an ACK back.
\item The ACK arrival tells the sender when the next packet in the same window should be released.
\item The next round uses the same timing pattern again.
\end{enumerate}
When the loop stabilizes, each flow effectively reuses the same slot positions from one RTT frame to the next. That is the mechanism behind the zero queue size. Formally, using a proof that is very similar to the previous theorem, we obtain:
\begin{theorem}
    Using CAPS in a dumbbell topology with heterogeneous RTTs, the bottleneck queue size converges to zero within an RTT.   
\end{theorem}

\begin{figure}
\centering
\begin{subfigure}{\columnwidth}
  \centering
  \includegraphics[width=0.9\columnwidth]{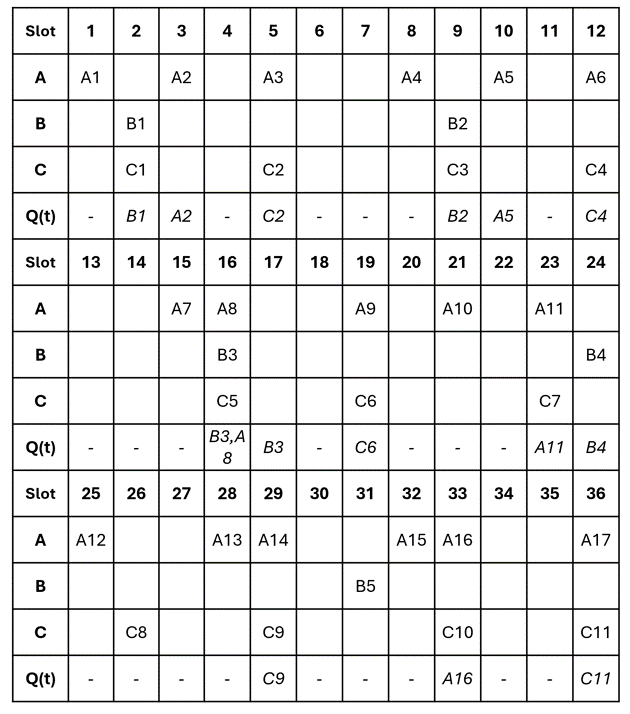}
  \caption{Baseline (no alignment): phase drift causes collisions.}
  \label{fig:ex2-base}
\end{subfigure}
\vspace{0.5em}
\begin{subfigure}{\columnwidth}
  \centering
  \includegraphics[width=0.9\columnwidth]{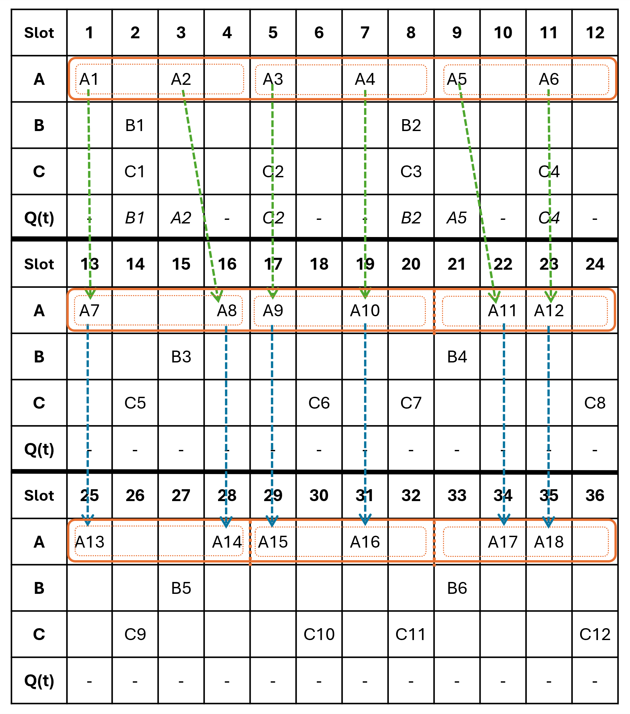}
  \caption{\name ($T_{\max}$ alignment): phase lock yields $100\%$ utilization with zero queue.}
  \label{fig:ex2-name}
\end{subfigure}
\caption{Heterogeneous-RTT example. Flow~A: $W_m{=}6$, RTT${=}4$. Flow~B: $W_m{=}2$, RTT${=}6$. Flow~C: $W_m{=}4$, RTT${=}3$. $T_{\max}{=}12$, capacity${=}1$~pkt/slot.}
\label{fig:example2}
\end{figure}

\paragraph{Example: Heterogeneous RTTs with and without \name}
\Cref{fig:example2} traces three flows in a network with $T_{\max} = 12$, capacity $C=1$\,pkt/slot, and $100\%$ total link rate). For better intuition, we assume initial window pacing rather than a single burst.
With standard unmodified ACK clocking (\cref{fig:ex2-base}), phase drift causes collisions and queueing. 
With \name (Fig.~\ref{fig:ex2-name}), the $T_{\max} - T_i$ correction phase-locks all flows: by the second frame, $Q(t) = 0$ at $100\%$ utilization. The green dashed arrows show flow~A's ACK-to-release timing after the first RTT; the blue dashed arrows show how it later stays periodic.

\mypar{General Topology} A general topology complicates things in two ways. First, it may have several bottlenecks, both on ACKs and on packets, complicating timing issues. Thus, we do not attempt to zero queue sizes anymore, just to reduce them. Second, ACKs and packets coexist on links. Thus, we need to take into account ACK rates and ensure that the sum of packet and ACK link rates stays below  capacity. Since timing is important, we also want to reduce delays for ACKs. To do so, we assume that at each switch, ACKS and packets are queued in separate FIFO queues, and ACKs have strict priority. This is easily implemented in all modern switches (\eg by using strict priority Queue 7 on Cisco Silicon One).

\mypar{Flow changes} When a flow arrives or departs, the CCA rate allocator (\eg S-PERC) recomputes feasible rates and updates each sender's window $W^m_i$. We then apply \name without change. \name re-locks phases within one $T_{\max}$ frame, after which queues return to zero or near-zero occupancy.

\section{Evaluation}

In this section, we evaluate the above results, starting with a dumbbell topology then generalizing to a datacenter fat-tree topology.

\mypar{Settings}
We use a discrete-event simulator built on SimPy. We always assume 4K-byte packets and 64-byte ACKs. In all simulations, each flow starts after some random waiting time to avoid artificial synchronization, and simulations run long enough to reach steady state after the initial transient. The fixed feasible rates are assigned by using S-PERC as the preliminary step.

\mypars{Topologies} In the dumbbell topology, links have 100-Gbps link capacities and link delays are drawn from a truncated Gaussian (clipped to positive values) with mean $100\,\mu$s. In the unidirectional case, we run 100 equal-rate flows at $99\%$ aggregate load. In the bidirectional case, the 100 flows are split equally (50 in each direction) and we use a $96\%$ aggregate packet load (to also allow for ACK load).

In the datacenter topology, we assume an 8-ToR two-level fat-tree with 4~core switches and 4~hosts per ToR (32~hosts total), 1,600\,Gbps links, host-to-ToR average propagation delays of $0.5\,\mu\text{s}$, and ToR-to-core average propagation delays of $2\,\mu\text{s}$. Again, exact delays are drawn from a Gaussian distribution with standard deviation $16\%$ of the mean.

\mypars{Baseline} We compare CAPS against a baseline algorithm that uses ACK clocking, as well as a baseline augmented with IWP.

\begin{figure*}[h]
\centering
\begin{subfigure}[]{\textwidth}
  \centering
  \includegraphics[width=0.35\textwidth]{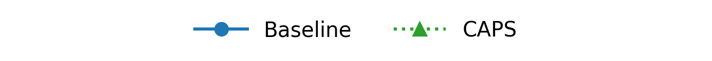}
\end{subfigure}
\begin{subfigure}[b]{0.32\textwidth}
    \includegraphics[width=\textwidth]{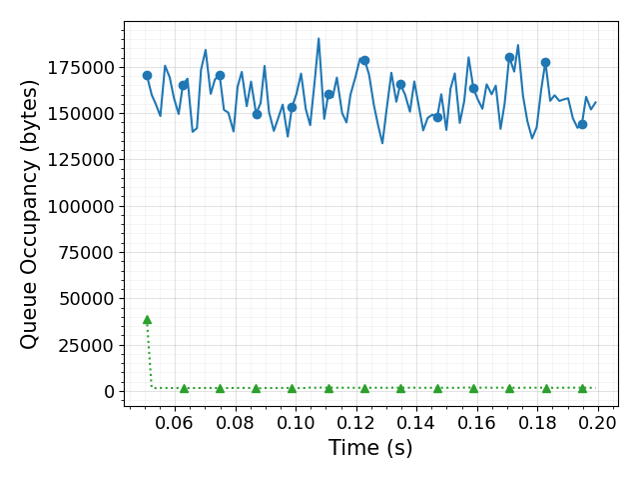}
    \caption{Unidirectional}
    \label{fig:uni-queue}
\end{subfigure}%
\hfill
\begin{subfigure}[b]{0.32\textwidth}
    \includegraphics[width=\textwidth]{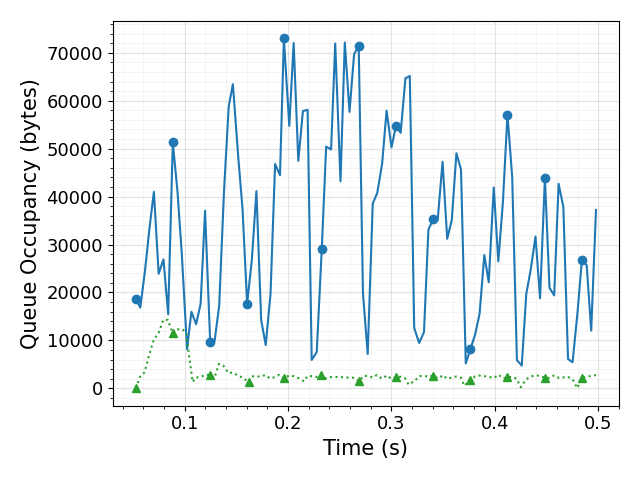}
    \caption{Bidirectional, forward}
    \label{fig:bi-fwd-queue}
\end{subfigure}%
\hfill
\begin{subfigure}[b]{0.32\textwidth}
    \includegraphics[width=\textwidth]{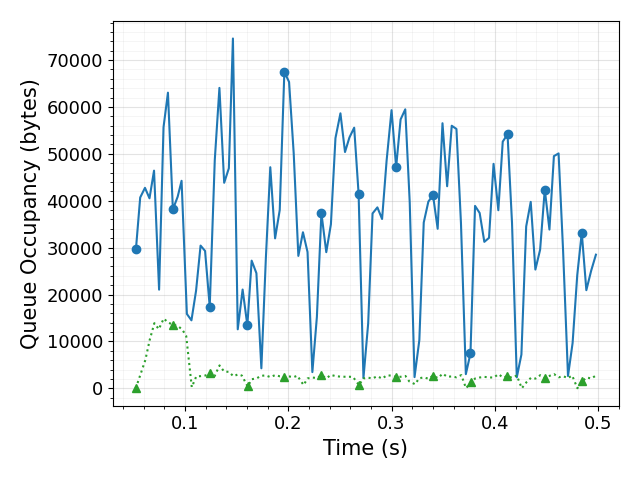}
    \caption{Bidirectional, reverse}
    \label{fig:bi-rev-queue}
\end{subfigure}
\caption{Dumbbell bottleneck queue occupancy in bytes as function of time with heterogeneous RTTs. (a) unidirectional, ~100 flows, $99\%$ load. (b,c) bidirectional, ~50 flows/direction, $96\%$ load.}
\label{fig:dumbbell-results}
\end{figure*}

\mypar{Dumbbell} \Cref{fig:dumbbell-results} illustrates the results of the dumbbell evaluations. \Cref{fig:uni-queue} shows how in the unidirectional case, after the initial transient (first $50\,\text{ms}$), the baseline accumulates a mean bottleneck occupancy of some $158\,\text{KB}$ (${\approx}40$ packets) with a peak near $200\,\text{KB}$ (${\approx}50$ packets). \textit{\name quickly reduces the mean occupancy to zero}, \ie every packet arriving at the bottleneck finds the link idle and is serialized immediately, while maintaining the same $99\%$ link utilization.

\Cref{fig:bi-fwd-queue,fig:bi-rev-queue} illustrate the bidirectional case.
After transient, the baseline sustains a mean forward queue of $34\,\text{KB}$ (${\approx}9$ packets) with peaks up to $76\,\text{KB}$ (${\approx}19$ packets), with similar numbers on the reverse path. \name does not yield a zero queue size anymore, as expected, but reduces the mean forward occupancy to $3.3\,\text{KB}$ (${\approx}1$ packet) with peaks near $16\,\text{KB}$ (${\approx}4$ packets), and the reverse direction is symmetric at $3.5\,\text{KB}$ mean. Both directions show an approximately $10\times$ reduction in mean occupancy compared to the baseline. The total throughput in both algorithms was near identical.

\mypar{Fat-tree topology}
We now run evaluations in a datacenter fat-tree topology. Each experiment compares three configurations: (i)~a simplified baseline with no alignment correction and no IWP, (ii)~the baseline with IWP added, and (iii)~\name with RTT$_{\max}$ alignment, split data/ACK queues, and IWP. In all cases, per-flow congestion windows are set by a max-min fair-rate allocator. The metric plotted is the maximum instantaneous queue occupancy across all switch links in the entire network, sampled at regular intervals. All plots begin slightly after simulation start (slightly over one RTT$_{\max}$) to exclude the initial burst caused by the first round of transmissions, which produces a large transient peak (predominantly in the unpaced baseline) that would compress the steady-state detail of interest.

\mypars{All-to-one (incast)}
All sources send to a single destination host, each flow split into 8~subflows (224~flows total) with randomized routing and a utilization booster bringing the bottleneck to $96\%$ load across all configurations. 

\begin{figure*}[h]
\centering
\begin{subfigure}[]{\textwidth}
  \centering
  \includegraphics[width=0.4\textwidth]{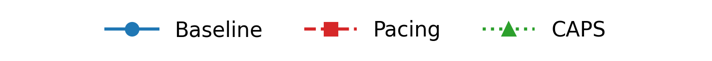}
\end{subfigure}
\begin{subfigure}[b]{0.32\textwidth}
    \includegraphics[width=\textwidth]{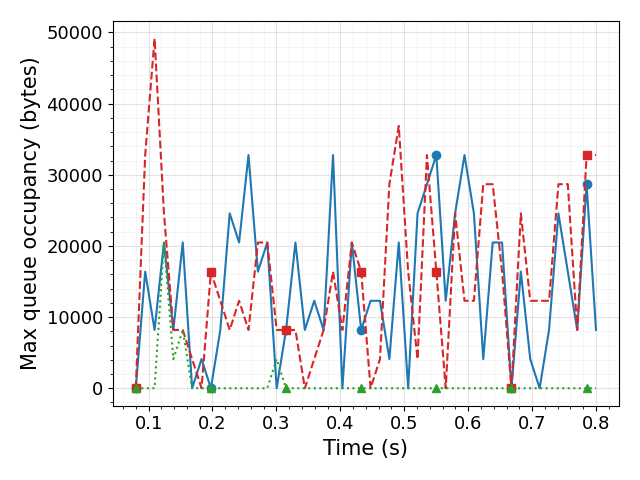}
    \caption{All-to-One (Incast)}
    \label{fig:allto1-queue}
\end{subfigure}%
\hfill
\begin{subfigure}[b]{0.32\textwidth}
    \includegraphics[width=\textwidth]{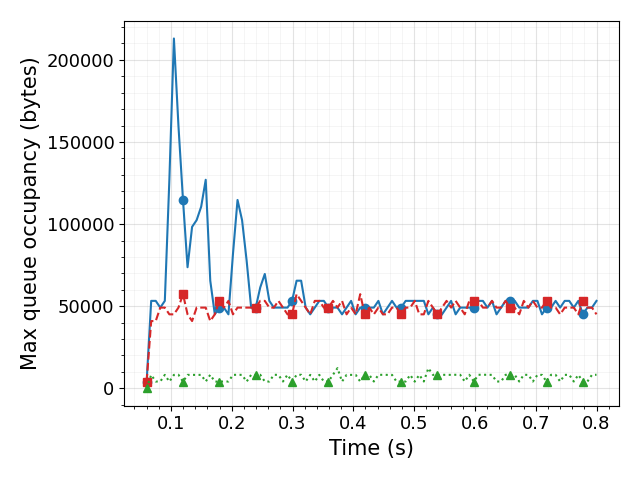}
    \caption{Permutation}
    \label{fig:derange-queue}
\end{subfigure}%
\hfill
\begin{subfigure}[b]{0.32\textwidth}
    \includegraphics[width=\textwidth]{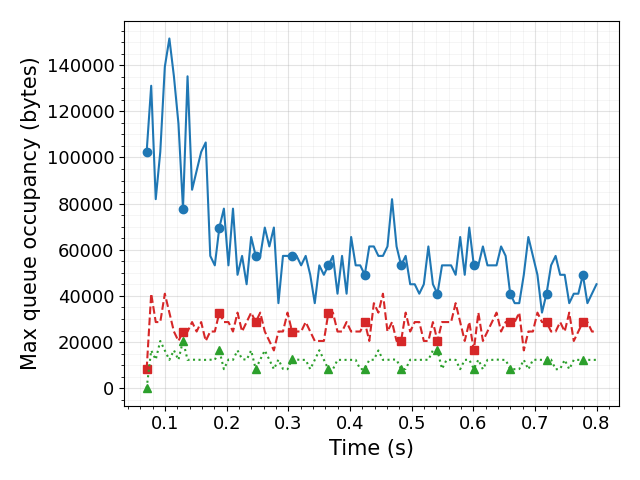}
    \caption{All-to-All}
    \label{fig:alltoall-queue}
\end{subfigure}
\caption{Fat-tree queue occupancy in bytes as function of time. (a)~bottleneck link queue, 224 flows, single bottleneck, $96\%$ load. (b)~max queue across all links, 32 permutation flows, multiple bottlenecks. (c)~max queue across all links, 896 flows, full contention.}
\label{fig:fattree-results}
\end{figure*}

\Cref{fig:allto1-queue} shows the last-hop bottleneck queue. In steady state, \name flattens it to ${\approx}1.5\,\text{KB}$, a $10\times$ reduction: even with 224~flows arriving through 4~core switches, phase lock keeps the single bottleneck nearly empty.

\mypars{Permutation traffic}
A random seeded permutation matches each of the 32~hosts to a distinct destination on a different ToR (32~flows), creating multiple simultaneous bottlenecks with no single dominant link. 

\Cref{fig:derange-queue} plots the network-wide maximum queue. IWP alone does not reduce steady-state buildup ($47$--$53\,\text{KB}$). \name reduces the worst-case queue to ${\approx}5$--$8\,\text{KB}$, a $7\times$ reduction,  demonstrating that per-flow phase alignment stabilizes timing independently at whichever link constrains each flow.

\mypars{All-to-all}
Every source sends to every host on every other ToR (896~flows), saturating all uplinks, core links, and downlinks simultaneously. 

\Cref{fig:alltoall-queue} shows the network-wide maximum queue. IWP reduces both the transient (baseline's ${\approx}310\text{KB}$ to ${\approx}35\,\text{KB}$ peak) and the steady state (baseline's $50$--$65$ to ${\approx}25$--$35\,\text{KB}$). \name further reduces the steady state queue to ${\approx}10$--$12\,\text{KB}$, $5\times$ from baseline and $2.5\times$ from IWP. The higher residual compared to permutation is expected: with dozens of simultaneously saturated links, cross-path interference limits phase-lock precision. Nevertheless, the reduction is substantial and consistent throughout the run.

\mypar{Result summary}
Table~\ref{tab:fattree-summary} summarizes the steady-state queue behavior across the three fat-tree scenarios.

\begin{table}
\centering
\caption{Average of the maximum queue occupancy across fat-tree experiments (after initial transient).}
\label{tab:fattree-summary}
\begin{tabular}{lccc}
\toprule
Scenario & Baseline & Baseline+IWP & \name \\
\midrule
All-to-One & $15\,\text{KB}$ & $14\,\text{KB}$ & $1.5\,\text{KB}$ \\
Permutation & $50\,\text{KB}$ & $50\,\text{KB}$ & $7\,\text{KB}$ \\
All-to-All & $55\,\text{KB}$ & $28\,\text{KB}$ & $11\,\text{KB}$ \\
\bottomrule
\end{tabular}
\end{table}

\section{Discussion}

\T{The missing primitive is timing, not rate.}
Current stacks control \emph{how much} to send and \emph{how to react} to congestion. \name adds a third dimension: \emph{when} to send. Two flows with identical rates can produce very different queue behavior depending on phase alignment. Treating release timing as a first-class control variable, alongside window and rate, makes a class of persistent queues avoidable by construction.

\T{Combining with existing transports.}
Swift and HPCC keep standing queues at a target (e.g., a few packets) by reacting to delay or INT signals, but their feedback loops cannot prevent persistent phase collisions: they only react after the queue has formed. We thought at first that \name is complementary: it could run beneath any such CCA, consuming the residual timing error that the feedback loop cannot eliminate. However, we found that Swift and HPCC \textit{need} to have queues to be able to react. It is an open problem to classify which CCAs coexist well with CAPS, and how to adapt the non-coexisting ones to obtain similar properties.

\T{Implications for buffer sizing.}
The small buffer-sizing rule allocates $B = \text{BDP}/\sqrt{N}$ on the premise that statistical multiplexing of $N$ flows still requires meaningful buffering to absorb rate fluctuations~\cite{buffer-sizing}. CAPS's results suggest this premise may over-provision for workloads where phase alignment is feasible. In the dumbbell with 100 flows at $99\%$ load, the standing queue drops to zero, far below what any BDP-based formula would prescribe. In the fat-tree with 896~flows, the worst-case queue falls to $2$--$3$~packets across the entire fabric. If phase-locked transport becomes practical at scale, buffer-sizing rules may need to distinguish between workloads that can be phase-aligned (long-lived, predictable) and those that cannot (short bursts, unpredictable arrivals), rather than applying a single statistical model to all traffic.

\T{Relevance to AI/ML collectives.}
GPU training clusters run synchronized AllReduce/AllGather~\cite{ml-collectives} where tail latency on any flow delays the entire collective~\cite{tail-at-scale}. \name's assumptions (stable rates, known RTTs, predictable patterns) are a natural fit for this workload, and the $5\times$ queue reduction in our all-to-all experiment suggests potential benefit. Production collectives involve per-step bursts and RDMA semantics that require further validation.

\T{Limitations.}
\name assumes (1)~converged fixed rates, (2)~no link oversubscription and (3) fixed routes. It is a post-convergence mechanism: during rate transitions, queues build until re-locking completes (one $T_{\max}$ transient). It is best suited for long-lived, predictable workloads (bulk transfers, AI/ML collectives). Open challenges for future work include $T_{\max}$ estimation in production and behavior under dynamic routing. 

\section{Conclusion}

\name is a timing layer on top of fair-rate control. The fair-rate allocator decides how much each flow may send; \name decides \emph{when} each flow should send so that packets reuse the same bottleneck slots from one RTT frame to the next. In equal-RTT settings, plain ACK-clocking is sufficient. In heterogeneous-RTT settings, a simple RTT$_{\max}$ correction restores alignment. In bidirectional settings, split data/ACK queues with ACK priority keep the feedback signal reliable.

The dumbbell experiments confirmed the mechanism in its cleanest form: zero steady-state queue at $99\%$ utilization with 100 heterogeneous-RTT flows, and a $10\times$ reduction under bidirectional traffic at $96\%$ load. The fat-tree experiments extended the result to a realistic multi-bottleneck topology with up to 896 concurrent flows across three traffic patterns (incast, random permutation, and all-to-all). CAPS reduced worst-case queue occupancy anywhere in the network by $5$--$10\times$ without throughput loss, using the same algorithm and parameters as the dumbbell case.

To summarize, the central finding is that standing queues are a timing bug, not a throughput requirement. High utilization and near-empty queues are not in conflict; they coexist naturally once flows are phase-aligned to the bottleneck service rhythm. 

\bibliographystyle{ACM-Reference-Format}
\bibliography{mybib} 

\end{document}